\documentclass[preprint,eqsecnum,preprintnumbers,nofootinbib,byrevtex,prd,aps,showpacs,showkeys,groupedaddress,floatfix]{revtex4}
\usepackage{bm}
\usepackage{graphics}
\usepackage{graphicx}
\usepackage{epsfig}
\usepackage{amssymb}
\usepackage{amsmath}

\newcommand\ba{\begin{eqnarray}}
\newcommand\ea{\end{eqnarray}}

\begin{document}

\date{}
\title{The bound state  solutions of the $D$-dimensional
Schr\"{o}dinger equation for the Woods-Saxon potential}
\author{V.~H.~Badalov}\email{E-mail:badalovvatan@yahoo.com}

\affiliation{Institute for Physical Problems, Baku State University,
Z. Khalilov st. 23, AZ-1148, Baku, Azerbaijan}


\begin{abstract}
In this work, the analytical solutions of the $D$-dimensional
Schr\"{o}dinger  equation are studied in great detail for the
Wood-Saxon potential by taking advantage of the Pekeris
approximation.  Within a novel improved scheme to surmount
centrifugal term, the energy eigenvalues and corresponding radial
wave functions are found for any angular momentum case within the
context of the Nikiforov-Uvarov (NU) and Supersymmetric quantum
mechanics (SUSYQM) methods. In this way, based on these methods, the
same expressions are obtained for the energy eigenvalues, and the
expression of radial wave functions transformed each other is
demonstrated. In addition, a finite number energy spectrum depending
on the depth of the potential $V_{0}$, the radial $n_{r}$ and
orbital $l$ quantum numbers and parameters $D, a, R_{0}$ are defined
as well.
\end{abstract}


\maketitle

\section{\bf Introduction}

An analytical solution of the radial Schr\"{o}dinger equation with a
physical potential is of paramount importance in nonrelativistic
quantum physics since the wave function and its associated
eigenvalues contain all necessary information for full description
of a quantum system. Along the years, there was a huge amount of
research effort to solve exactly the radial Schr\"{o}dinger equation
for all values of $n_{r}$ and $l$ quantum numbers, but it could only
be possible for a few specific potentials. In this way, there are
several established analytical methods, including Polynomial
solution ~\cite{Landau,Greiner,Flugge}, Nikiforov - Uvarov method
(NU) ~\cite{Nikiforov}, Supersymmetric quantum mechanics method
(SUSYQM) ~\cite{Cooper,Morales}, and Asymptotic iteration method
(AIM) ~\cite{Ciftci1,Ciftci2,Ciftci3,Bayrak1,Bayrak2,
Bayrak3,Ciftci4}, to solve the radial Schr\"{o}dinger equation
exactly or quasi-exactly for $l\neq0$ within these potentials.
G.Levai et al. suggested a simple method for the proposed potentials
for which the Schr\"{o}dinger equation can be solved exactly with
special functions ~\cite{Levai} and presented relationship between
the introduced formalism and SUSYQM ~\cite{Cooper}. Furthermore, in
order to solve the Schr\"{o}dinger equation applicable to problems
of nonperturbative nature, P.Amore et al. introduced a novel method
~\cite{Amore}. Thus, this method was applied to calculate the
energies and wave functions of the ground and first excited state of
the quantum anharmonic potential. It is well known that the
Woods-Saxon potential ~\cite{Woods} which we consider in the present
work is one of the most realistic short-range potentials in physics.
This potential plays a vital role in many branches of physics such
as nuclear and particle, atomic, condensed matter, molecular and
chemical physics. In fact, this potential cannot be solved exactly
without using any approximation for $l\neq0$ yet. However,
S.Fl\"{u}gge gave an exact expression for the wave function and
suggested a graphical method for the energy eigenvalues at $l=0$
~\cite {Flugge}.

In recent years, the NU~\cite {Nikiforov} and SUSYQM
~\cite{Cooper,Morales} methods with various approximations have been
proposed for solving the Schr\"{o}dinger equation analytically. Many
papers show the power and simplicity of both of these methods in
solving central and noncentral potentials ~\cite
{Badalov1,Badalov2,Badalov3,Badalov4, Ahmadov1,Ahmadov2,
Ikhdair1,Ikhdair2,Ikhdair3,Ikhdair4,Badalov5}. The NU method is
based on solving the second-order linear differential equation by
reducing to a generalized equation of hypergeometric-type which is a
second-order type homogeneous differential equation with polynomials
coefficients of degree not exceeding the corresponding order of
differentiation, and SUSYQM method allows one to determine
eigenstates of known analytically solvable potentials using
algebraic operator formalism without ever having to solve the
Schr\"{o}dinger differential equation by standard series technique.
Hence it would be interesting and important to solve the
nonrelativistic radial Schr\"{o}dinger equation for Woods-Saxon
potential for $l\ne 0$, since it has been extensively utilized to
describe the bound and continuum states of the interacting systems.
In this way, one can need to obtain the energy eigenvalues and
corresponding eigenfunctions of the one particle problem within this
potential. The NU method was used by C.Berkdemir et al.~\cite
{Berkdemir} solved the radial Schr\"{o}dinger equation for the
generalized Woods-Saxon potential for $l=0$. However, in this work,
the authors made errors in application of the NU method, and these
errors have led to incorrect results ~\cite {Editorial}. In the
following works ~\cite {Ikhdair5,Arda1,Arda2, Falaye,Ikhdair6},
authors made similar errors in application of the NU and AIM
methods.

In our previous works ~\cite {Badalov1,Badalov2,Badalov3,Badalov4},
for the first time, we have given the comprehensive information of
how to obtain analytically the exact energy eigenvalues and the
corresponding wave functions of the radial Schr\"{o}dinger and the
radial Klein-Gordon equations with Woods-Saxon potential via NU
method. In these works, the below approximation scheme - the Pekeris
approximation ~\cite {Pekeris} - was proposed for
$V_{l}(r)=\frac{\hbar^{2} \tilde{l}(\tilde{l}+1)}{2\mu r^{2} }$ the
centrifugal potential in any arbitrary $l$ state:

\begin{equation}
\frac{1}{r^{2} } =\frac{1}{R_{0}^{2} } \left(C_{0} +\frac{C_{1}
}{1+e^{\frac{r-R_{0} }{a} } } +\frac{C_{2}
}{\left(1+e^{\frac{r-R_{0} }{a} } \right)^{2} } \right)\, ,
\end{equation}
where $C_{0}, C_{1}, C_{2}$  quantities dependent on $R_{0}, a$
specific potential parameters were defined by comparing both sides
of Eq.(1.1) expression expanding in the Taylor series around the
point $r=R_{0}$. Furthermore, the D-dimensional Schr\"{o}dinger
equation with Woods-Saxon potential was solved within the context of
the NU, AIM, and SUSYQM methods through the Pekeris approximation to
the centrifugal potential, and the energy eigenvalues and
corresponding radial wave functions are found for any arbitrary
state as well ~\cite {Badalov5}.

In this study, the exact nonzero angular momentum solutions to
$D$-dimensional Schr\"{o}dinger equation with Woods-Saxon potential
are presented within the context of the NU and SUSYQM methods. Since
a nucleon in the interior of the nucleus feels a different potential
than that near the surface of the nucleus, the study is extended
there by using a new improved approximate scheme to deal with the
centrifugal term, i.e., the Pekeris approximation is applied to the
centrifugal potential $V_{l}(r)$  based on the settings $C_{0},
C_{1}, C_{2}$ quantities which are determined around the point
$r=r_{min}$ of the effective $l$ dependent potential $V_{eff}(r)$.
Thus, the best part of this research work is that the same
expressions are obtained for the energy eigenvalues and
corresponding eigenfunctions in various values of $n_{r}$  and $l$
quantum numbers by taking advantage of the NU and SUSYQM methods.

This paper is structured as follows: After this introduction, the
detailed description of the  $D$-dimensional SE with the Woods-Saxon
potential is given in Section II. Then in Section III, the
impletemetion of  NU method to $D$-dimensional SE is stated. In
Section IV, the application of SUSYQM method to  $D$-dimensional SE
is presented. Next, the results and discussion are presented in
Section V. Finally, the paper is concluded with brief summary in
Section VI. Appendies A and B contain an overview of NU and SUSYQM
methods respectively.

\section{\bf  The $D$-dimensional Schr\"{o}dinger  equation with the Woods-Saxon potentials}

Woods and Saxon proposed a potential to describe the distribution of
20 MeV protons on the heavy nuclei, such as platinum or nickel [16].
The spherical Woods-Saxon potential that was widely used to study
the nuclear structure within the shell model has received
significant attention in nuclear mean field model. The spherical
standard Woods-Saxon potential ~\cite {Woods} is defined by

\begin{equation}
V(r)=-\frac{V_{0} }{1+e^{\frac{r-R_{0} }{a} } } \, \, \, \, \, \, \,
\, (a<<R_{0} ).
\end{equation}

This potential was also considered for description of interaction of
a neutron with a heavy nucleus. The parameter $R_{0}$ represents the
width of the nuclear radius; the parameter $a$ characterizes
thickness of the superficial layer inside which the potential falls
from value $V=0$ outside of a nucleus up to value $V=-V_{0}$ inside
a nucleus. At $a=0$, one gets the simple potential well with jump of
potential on the surface of a nucleus.

Using $D$-dimensional $(D\ge {\rm 2)}$ polar coordinates with polar
variable $r$ (hyperradius) and angular variables $r,\, \, \theta
_{1} \, ,\, \theta _{2} \, ,\, \ldots ,\, \theta _{D-2} \, ,\, \phi
$ (hyperangles), the Laplasian operator in polar coordinates $r,\,
\, \theta _{1} \, ,\, \theta _{2} \, ,\, \ldots ,\, \theta _{D-2} \,
,\, \phi $ of $R^{D}$ is

\begin{equation}
\nabla _{D}^{2} =r^{1-D} \frac{\partial }{\partial r} \left(r^{D-1}
\frac{\partial }{\partial r} \right)+\frac{\Lambda _{D}^{2} }{r^{2}
}  ,
\end{equation}
where $\Lambda _{D}^{2}$ is a partial differential operator on the
unit sphere $S^{D-1}$ (Laplace-Beltrami operator or grand orbital
operator or hyperangular momentum operator) defined analogously to a
three-dimensional ($3D$) angular momentum by Avery ~\cite {Avery}.

The $D$-dimensional Schr\"{o}dinger equation with spherically
symmetric potential $V(r)$ is of the form ~\cite {Avery}

\begin{equation}
\left(-\frac{\hbar ^{2} }{2\mu } \nabla _{D}^{2} +V(r)-E_{n_{r}l}
\right)\psi _{n_{r}lm} (r,\Omega _{D} )=0\, ,
\end{equation}
where $\mu $ is the reduced mass, $\hbar $ is the Planck's constant
and

\begin{equation}
\psi _{n_{r}lm} (r,\Omega _{D} )=R_{n_{r}l}(r)Y_{lm} (\Omega _{D}
)\, .
\end{equation}
The Laplasian operator divides into a hyper-radial part $r^{1-D}
\frac{\partial }{\partial r} \left(r^{D-1} \frac{\partial }{\partial
r} \right)$ and an angular part $\frac{\Lambda _{D}^{2} }{r^{2} }
=-\frac{\hat{L}_{D}^{2} }{\hbar ^{2} r^{2} } $ i.e.

\begin{equation}
\nabla _{D}^{2} =r^{1-D} \frac{\partial }{\partial r} \left(r^{D-1}
\frac{\partial }{\partial r} \right)-\frac{\hat{L}_{D}^{2} }{\hbar
^{2} r^{2} },
\end{equation}
where $\hat{L}_{D} $ is the grand orbital angular momentum operator.
The eigenfunctions of $\hat{L}_{D}^{2} $ are the hyper-spherical
harmonics

\begin{equation}
\hat{L}_{D}^{2} Y_{lm} (\Omega _{D} )\, =\hbar ^{2} l(l+D-2)Y_{lm}
(\Omega _{D} )\, ,
\end{equation}
where $l$ is the angular momentum quantum number.

After substituting Eqs.(2.4) - (2.6) into (2.3) and using $\psi
_{n_{r}lm} (r,\Omega _{D} )$ as the eigenfunction of
$\hat{L}_{D}^{2} $ with eigenvalue $\hbar ^{2} l(l+D-2)$, we obtain
an equation known as the hyper-radial Schr\"{o}dinger equation with
Woods-Saxon potential

\begin{equation}
\frac{d^{2} R_{n_{r}l} (r)}{dr^{2} } +\frac{D-1}{r}
\frac{dR_{n_{r}l} (r)}{dr} +\frac{2\mu }{\hbar ^{2} } \left[E_{nl}
-V(r)\, \, -\frac{\hbar ^{2} l(l+D-2)}{2\mu r^{2} }
\right]R_{n_{r}l} (r)=0\, ,\, \, (0\le r<\infty ).
\end{equation}
Introducing a new function $u_{n_{r}l} (r)=r^{\, \frac{D-1}{2} }
R_{n_{r}l} (r)$, Eq.(2.7) reduces

\begin{equation}
\frac{d^{2} u_{n_{r}l} (r)}{dr^{2} } +\frac{2\mu }{\hbar ^{2}}
\left[E_{n_{r}l} -V(r)\, \, \, -\frac{\hbar ^{2}
\left(l+\frac{D-1}{2} \right)\left(l+\frac{D-3}{2} \right)}{2\mu
r^{2} } \right]u_{n_{r}l} (r)=0\, ,
\end{equation}
and introducing a new parametr $\tilde{l}=l+\frac{D-3}{2} $ ,
Eq.(2.8) takes the form

\begin{equation}
\frac{d^{2} u_{n_{r}l} (r)}{dr^{2} } +\frac{2\mu }{\hbar ^{2} }
\left[E_{n_{r}l} -V_{eff} (r)\right]u_{n_{r}l} (r)=0,
\end{equation}
where $V_{eff} (r)$ is effective potential, i.e.

\begin{equation}
V_{eff}(r)=V(r)+\frac{\hbar ^{2} \tilde{l}(\tilde{l}+1)}{2\mu r^{2}
}   .
\end{equation}

Equation (2.9) has the same form as the equation for a particle in
one dimension, except for two important differences. First, there is
a repulsive effective potential proportional to the eigenvalue of
$\hbar ^{2} \tilde{l}(\tilde{l}+1)$. Second, the radial function
must satisfy the boundary conditions $u(0)=0$ and $u(\infty )=0$.

It is well-known that the Schr\"{o}dinger equation cannot be solved
exactly for this potential at the value $l\ne 0$ by using the
standard methods as SUSY and NU. From Eq.(2.10), it is seen that the
effective potential is combination of the exponential and inverse
square potentials which cannot be solved analytically. That is why,
in order to solve this problem we can take the most widely used and
convenient for our purposes Pekeris approximation. This
approximation is based on the expansion the series for exponential
cases depending on the internuclear distance of the centrifugal
barrier, and there the terms up to second-order are considering.

After introducing the new variable  $x=\frac{r-R_{0} }{R_{0} }$ or
$r=R_{0} (1+x)$, the effective Woods-Saxon potential as following
form:
\begin{equation}
V_{eff} (r)=-\frac{V_{0} }{1+e^{\alpha
x}}+\frac{\tilde{\delta}}{(1+x)^{2}},
\end{equation}
where $\alpha=\frac{R_{0} }{a}$   and $\tilde{\delta }=\frac{\hbar
^{2} \tilde{l}(\tilde{l}+1)}{2\mu R_{0}^{2} } $. The extreme point
of the effective potential $V_{eff} (r)$  is defined by the
following equation
\begin{equation}
\frac{\alpha V_{0} \, e^{\alpha x} }{\left(1+e^{\alpha x}
\right)^{2} } \, \, = \frac{2\tilde{\delta}}{(1+x)^{3}}.
\end{equation}
Since the solution of Eq.(2.12) depends on orbital $l$ quantum
numbers, $x=x_{\min }=x_{l}$ ($r=r_{\min }=r_{l}$).

Let us expand centrifugal potential $V_{l} (r)$ in Taylor series
around the point of $x=x_{l} $ \;($r=r_{l}$) satisfied the
transcendent Eq. (2.12):

\begin{equation}
\begin{array}{l} {V_{l} (r)=\frac{\hbar ^{2} \tilde{l}(\tilde{l}+1)}{2\mu r^{2}
} =\frac{\hbar ^{2} \tilde{l}(\tilde{l}+1)}{2\mu R_{0}^{2} } \cdot
\frac{\tilde{\delta }}{(1+x)^{2} } =\tilde{\delta
}\left[\frac{1}{(1+x_{l} )^{2} } -\frac{2}{(1+x_{l} )^{3} } \cdot
(x-x_{l} )+\right. } \\ {\left. \, \, \, \, \, \, \, \, \, \, \, \,
\, \, \, \, \, \, \, \, \, \, \, \, \, \, \, \, \, \, \, \, \, \, \,
\, \, \, \, \, \, \, \, \, \, \, \, \, \, \, \, \, \, \, \, \, \, \,
\, \, \, \, \, \, \, \, \, \, \, +\, \frac{3}{(1+x_{l} )^{3} } \cdot
(x-x_{l} )^{2} +o((x-x_{l} )^{3} )\right]} \end{array}.
\end{equation}
According to the Pekeris approximation, $V_{l}(r)$ takes the form
~\cite {Badalov1,Badalov2,Badalov3,Badalov4}

\begin{equation}
\tilde{V}_{l} (r)=\tilde{\delta }\left(C_{0} +\frac{C_{1}
}{1+e^{\alpha x} } +\frac{C_{2} }{(1+e^{\alpha x} )^{2} } \right).
\end{equation}
Let us expand the potential $\tilde{V}_{l} (r)$ in the Taylor series
around the point of $x=x_{l} $ \;($r=r_{l} $) :

\begin{equation}
\begin{array}{l} {\tilde{V}_{l} (x)=\tilde{\delta }\left[C_{0} +\frac{C_{1} }{1+e^{\alpha x_{l} }
} +\frac{C_{2} }{(1+e^{\alpha x_{l} } )^{2} } -\left(\frac{\alpha
C_{1} \, e^{\alpha x_{l} } }{(1+e^{\alpha x_{l} } )^{2} }
+\frac{2\alpha C_{2} \, e^{\alpha x_{l} } }{(1+e^{\alpha x_{l} }
)^{3} } \right)(x-x_{l} )+ \right. } \\ \left.{\, \, \, \, \, \, \,
\, \, \, \, \, -\left(\frac{\alpha ^{2} C_{1} \, e^{\alpha x_{l} }
(1-e^{\alpha x_{l} } )}{2(1+e^{\alpha x_{l} } )^{3} } +\frac{\alpha
^{2} C_{2} \, e^{\alpha x_{l} } (1-2e^{\alpha x_{l} }
)}{(1+e^{\alpha x_{l} } )^{4} } \right)(x-x_{l} )^{2} +o((x-x_{l}
)^{3} )} \right]  \end{array}.
\end{equation}

In order to define the constants $C_{0} \, ,\, C_{1} $ and $C_{2}$,
we compare the compatible degrees of same order of $x$ in Eqs.(2.13)
and (2.15), and obtain the following algebraic system equations:

\begin{equation}
\left\{\begin{array}{l} {C_{0} +\frac{C_{1} }{1+e^{\alpha x_{l} }
}+\frac{C_{2} }{(1+e^{\alpha x_{l} } )^{2} } =\frac{1}{(1+x_{l}
)^{2} } } \\ {\frac{\alpha C_{1} e^{\alpha x_{l} } }{(1+e^{\alpha
x_{l} } )^{2} } +\frac{2\alpha C_{2} e^{\alpha x_{l} }
}{(1+e^{\alpha x_{l} } )^{3} } =\frac{2}{(1+x_{l} )^{3} } } \\
{\frac{\alpha ^{2} C_{1} \, e^{\alpha x_{l} } (1-e^{\alpha x_{l} }
)}{2(1+e^{\alpha x_{l} } )^{3} } +\frac{\alpha ^{2} C_{2} \,
e^{\alpha x_{l} } (1-2e^{\alpha x_{l} } )}{(1+e^{\alpha x_{l} }
)^{4} } =-\frac{3}{(1+x_{l} )^{4} } } \end{array}\right. .
\end{equation}

From the  solution of Eq.(2.16) algebraic system equations, for
$C_{0} \, ,\, C_{1} $ and $C_{2}$ constants, we get the following
relations:

\begin{equation}
\left\{\begin{array}{l} {C_{0} =\frac{1}{(1+x_{l} )^{2} }
+\frac{(1+e^{\alpha x_{l} } )^{2} }{\alpha e^{\alpha x_{l} }(1+x_{l}
)^{3} } \left[\frac{e^{-\alpha x_{l} } -3}{1+e^{\alpha x_{l} } }
+\frac{3e^{-\alpha x_{l} } }{\alpha (1+x_{l} )} \right]} \\ {C_{1}
=\frac{2(1+e^{\alpha x_{l} } )^{2} }{\alpha e^{\alpha x_{l} }
(1+x_{l} )^{3} } \left[2-e^{-\alpha x_{l} } -\frac{3(1+e^{-\alpha
x_{l} } )}{\alpha (1+x_{l} )} \right]} \\ {C_{2} =\frac{(1+e^{\alpha
x_{l} } )^{3} }{\alpha e^{\alpha x_{l} } (1+x_{l} )^{3} }
\left[e^{-\alpha x_{l} } -1+\frac{3(1+e^{-\alpha x_{l} } )}{\alpha
(1+x_{l} )} \right]} \end{array}\right. .
\end{equation}

After Pekeris approximation, the effective potential as the
following form:

\begin{equation}
\tilde{V}_{eff} (r)=K_{0} -\frac{K_{1} }{1+e^{\frac{r-R_{0} }{a} }
}+\frac{K_{2} }{\left(1+e^{\frac{r-R_{0} }{a} } \right)^{2} } ,
\end{equation}
where $K_{0}=\tilde{\delta }C_{0}, K_{1}=V_{o}-\tilde{\delta }C_{1},
K_{2}=\tilde{\delta }C_{2} $ , i.e.,
\begin{equation}
\left\{\begin{array}{l} {K_{0} =\frac{\tilde{\delta }}{(1+x_{l}
)^{2} }+\frac{(1+e^{\alpha x_{l} } )^{2} \tilde{\delta }}{\alpha
e^{\alpha x_{l} } (1+x_{l} )^{3} } \left[\frac{e^{-\alpha x_{l} }
-3}{1+e^{\alpha x_{l} } } +\frac{3e^{-\alpha x_{l} } }{\alpha
(1+x_{l} )} \right]} \\ {K_{1} =V_{0}+\frac{2(1+e^{\alpha x_{l} }
)^{2} \tilde{\delta }}{\alpha e^{\alpha x_{l} } (1+x_{l} )^{3} }
\left[e^{-\alpha x_{l} } -2+\frac{3(1+e^{-\alpha x_{l} } )}{\alpha
(1+x_{l} )} \right]} \\ {K_{2} =\frac{(1+e^{\alpha x_{l} } )^{3}
\tilde{\delta }}{\alpha e^{\alpha x_{l} } (1+x_{l} )^{3} }
\left[e^{-\alpha x_{l} } -1+\frac{3(1+e^{-\alpha x_{l} } )}{\alpha
(1+x_{l} )} \right]} . \end{array}\right.
\end{equation}

If we consider $x_{l} =0$ in Eq.(2.17) relations, the constants
$C_{0} \, ,\, C_{1} $ and $C_{2}$ can be written in a closed form as
~\cite {Badalov1,Badalov2,Badalov3,Badalov4}:

\[C_{0} \, =1-\frac{4}{\alpha } +\frac{12}{\alpha ^{2}},
 C_{1} \, =\frac{8}{\alpha }-\frac{48}{\alpha ^{2} }, C_{2} \, =\frac{48}{\alpha ^{2} }.\]

According to Eq.(2.12), Eq.(2.19) as the following form:

\begin{equation}
\left\{\begin{array}{l} {K_{0} =\frac{V_{0}}{2}\left[\frac{\alpha
(1+x_{l})}{4\cosh^{2}(\frac{\alpha x_{l}}{2})}+\frac{e^{-\alpha
x_{l}}-3}{1+e^{\alpha x_{l}
}} +\frac{3e^{-\alpha x_{l}}}{\alpha (1+x_{l})}\right]} \\
{K_{1}={V_{0}}\left[e^{-\alpha x_{l}}
-1+\frac{3(1+e^{-\alpha x_{l}})}{\alpha (1+x_{l})}\right]} \\
{K_{2} ={V_{0}}\left[\frac{6\cosh^{2}(\frac{\alpha x_{l}}{2})}{
\alpha (1+x_{l})}-\sinh (\alpha x_{l}) \right]} .
\end{array}\right.\frac{}{}
\end{equation}

It should be noted that this approximation preserves the original
form of the effective $l$ dependent potential and is valid only for
low vibrational energy cases.

Instead of solving the hyper-radial Schr\"{o}dinger equation for the
effective Woods-Saxon potential $V_{eff}(r)$ given by Eq.(2.11), we
now solve the hyper-radial Schr\"{o}dinger equation for the new
effective potential $\tilde{V}_{eff}(r)$ given by Eq.(2.18) obtained
using the Pekeris approximation. Having inserted this new effective
potential into Eq.(2.9), we obtain

\begin{equation}
\frac{d^{2} u_{n_{r}l} (r)}{dr^{2} } +\frac{2\mu }{\hbar ^{2} }
\left[E_{n_{r}l} -K_{0} +\frac{K_{1} }{1+e^{\frac{r-R_{0} }{a} }
}-\frac{K_{2} }{\left(1+e^{\frac{r-R_{0} }{a} } \right)^{2} }
\right]u_{n_{r}l} (r)=0.
\end{equation}

If we rewrite equation Eq.(2.21) by using a new variable of the form
$z=\left(1+e^{\frac{r-R_{0} }{a} } \right)^{-1} $, we obtain

\begin{equation}
z^{2} (1-z)^{2} u''(z)+z(1-z)(1-2z)u'(z)+\frac{2\mu a^{2} }{\hbar
^{2} } \left[E-K_{0} +K_{1} z-K_{2} z^{2}\right]\, u(z)=0\, \, ,\,
\, \, (0\le z\le 1).
\end{equation}

We use the following dimensionless notations

\begin{equation}
\varepsilon ^{2} =-\frac{2\mu \, a^{2} (E-K_{0} )}{\hbar ^{2} }
>0,\, \, \beta ^{2}=\frac{2\mu \, a^{2} K_{1} }{\hbar ^{2} } >0,\,
\, \gamma ^{2} =\frac{2\mu \, a^{2} K_{2} }{\hbar ^{2} } >0,
\end{equation}
we obtain

\begin{equation}
u''(z)+\frac{1-2z}{z(1-z)} u'(z)+\frac{-\varepsilon ^{2} +\beta ^{2}
z-\gamma ^{2} z^{2}}{\left(z(1-z)\right)^{2} } u(z)=0\, \, ,\, \, \,
(0\le z\le 1)
\end{equation}
with real $\varepsilon >0$ \,($E<0$) for bound states; $\beta $ and
$\gamma $ are real and positive.

\section{\bf Solution of the $D$-dimensional Schr\"{o}dinger equation by Nikiforov-Uvarov Method}

According to the NU-method from Eqs.(A.1) and (2.24), we obtain

\begin{equation}
\tilde{\tau }(z)=1-2z;\, \, \sigma (z)=z(1-z);\, \, \tilde{\sigma
}(z)=-\varepsilon ^{2} +\beta ^{2} z-\gamma ^{2} z^{2}    ,
\end{equation}
and the new function $\pi (z)$ is

\begin{equation}
\pi (z)=\pm \sqrt{\varepsilon ^{2} + (k-\beta ^{2})z-(k-\gamma ^{2}
)z^{2}} .
\end{equation}

The constant parameter $k$ can be found employing the condition that
the expression under the square root has a double zero, i.e., its
discriminant is equal to zero. So, there are two possible functions
for each $k$

\begin{equation}
\pi (z)=\pm \left\{\begin{array}{l} {\left(\varepsilon
-\sqrt{\varepsilon ^{2} -\beta ^{2} +\gamma ^{2} }
\right)z-\varepsilon \, \, ,\, \, \, {\rm for}\, \, \, k=\beta ^{2}
-2\varepsilon ^{2} +2\varepsilon \sqrt{\varepsilon ^{2} -\beta ^{2}
+\gamma ^{2} } \, ,} \\ {\left(\varepsilon +\sqrt{\varepsilon ^{2}
-\beta ^{2} +\gamma ^{2} } \right)z-\varepsilon \, \, ,\, \, \, {\rm
for}\, \, \, k=\beta ^{2} -2\varepsilon ^{2} -2\varepsilon
\sqrt{\varepsilon ^{2} -\beta ^{2} +\gamma ^{2} } \, .}
\end{array}\right.
\end{equation}

According to the NU-method, from the four possible forms of the
polynomial $\pi(z)$ we select the one for which the function $\tau
(z)$ has the negative derivative and root lies in the interval
$(0,\, \, 1)$. Therefore, the appropriate functions $\pi(z)$ and
$\tau (z)$ have the following forms

\begin{equation}
\pi (z)=\varepsilon -\left(\varepsilon +\sqrt{\varepsilon ^{2}
-\beta ^{2} +\gamma ^{2} } \right)\, z,
\end{equation}

\begin{equation}
\tau (z)=1+2\varepsilon -2\left(1+\varepsilon +\sqrt{\varepsilon
^{2} -\beta ^{2} +\gamma ^{2} } \right)\, z,
\end{equation}
and
\begin{equation}
k=\beta ^{2} -2\varepsilon ^{2} -2\varepsilon \sqrt{\varepsilon ^{2}
-\beta ^{2} +\gamma ^{2} } .
\end{equation}
Then, the constant $\lambda =k+\pi '(z)$ is written as
\begin{equation}
\lambda =\beta ^{2} -2\varepsilon ^{2} -2\varepsilon
\sqrt{\varepsilon ^{2} -\beta ^{2} +\gamma ^{2} }-\varepsilon
-\sqrt{\varepsilon ^{2} -\beta ^{2} +\gamma ^{2} } .
\end{equation}
An alternative definition of $\lambda _{n_{r} } $ (Eq.(A.9)) is
\begin{equation}
\lambda =\lambda _{n_{r} } =2\left(\varepsilon +\sqrt{\varepsilon
^{2} -\beta ^{2} +\gamma ^{2} } \right)\, n_{r} +n_{r} (n_{r} +1).
\end{equation}
Having compared Eq.(3.7) with Eq.(3.8)
\begin{equation}
\beta ^{2} -2\varepsilon ^{2} -2\varepsilon \sqrt{\varepsilon ^{2}
-\beta ^{2} +\gamma ^{2} } -\varepsilon -\sqrt{\varepsilon ^{2}
-\beta ^{2} +\gamma ^{2} } =2\left(\varepsilon +\sqrt{\varepsilon
^{2} -\beta ^{2} +\gamma ^{2} } \right)\, n_{r} +n_{r} (n_{r} +1),
\end{equation}
we obtain

\begin{equation}
\varepsilon +\sqrt{\varepsilon ^{2} -\beta ^{2} +\gamma ^{2} }
+n_{r} +\frac{1}{2} -\frac{\sqrt{1+4\gamma ^{2} } }{2} =0
\end{equation}
or

\begin{equation}
\varepsilon +\sqrt{\varepsilon ^{2} -\beta ^{2} +\gamma ^{2} }
-n'=0.
\end{equation}
Here

\begin{equation}
n'=-n_{r} +\frac{\sqrt{1+4\gamma ^{2} } -1}{2},
\end{equation}
and $n_{r}$ is the radial quantum number $(n_{r} =0,\, 1,\, 2,\,
...)$. From Eq.(3.11), we find

\begin{equation}
\varepsilon =\frac{1}{2} \left(n'+\frac{\beta ^{2} -\gamma ^{2}
}{n'} \right).
\end{equation}

From the bound states $-V_{0} <E<0$ and finite wavefunction, we get
$\varepsilon >0$ and $\sqrt{\varepsilon ^{2} -\beta ^{2} +\gamma
^{2} } >0$, i.e. $n'>0$ and $|\beta ^{2} -\gamma ^{2}|<n'^{2} $.
According to Eqs.(3.12) and (2.12) this relations can be recast into
the form:

\begin{equation}
0\le n_{r} <\frac{1}{2} \left(\sqrt{1+\frac{8\mu \, a^{2} K_{2}
}{\hbar ^{2} } } -1\right),
\end{equation}

\begin{equation}
V_{0} R_{0}^{3}\geq \frac{4\hbar ^{2} \tilde{l}(\tilde{l}+1)a}{\mu
\, } .
\end{equation}
Substituting the values of $\varepsilon ,\, \, \beta ,\, \gamma$ and
$n'$ into Eq.(3.13), one can find energy eigenvalues
$E^{(D)}_{n_{r}l}$

\begin{equation}
E^{(D)}_{n_{r}l}=K_{0} -\frac{K_{1} -K_{2} }{2} -\frac{\hbar ^{2}
}{32\mu \, a^{2}} \left(\sqrt{1+\frac{8\mu \, a^{2} K_{2}}{\hbar
^{2}}} -2n_{r} -1\right)^{2} -\frac{\frac{2\mu \, a^{2}}{\hbar ^{2}}
(K_{1} -K_{2} )^{2}}{\left(\sqrt{1+\frac{8\mu \, a^{2} K_{2}}{\hbar
^{2}}} -2n_{r} -1\right)^{2}}.
\end{equation}

If the conditions Eq.(3.14) and Eq.(3.15) are satisfied
simultaneously, the bound states exist. Thus, the energy spectrum
Eq.(3.16) is limited, i.e. we have only the finite number of energy
eigenvalues.

For very large $V_{0}$, the $l$-dependent effective potential has
the same form as the potential with $l=0$. When $D=3$, from
Eq.(3.14) is seen that if $l=0$, then one gets $n_{r}<0$. Hence, the
Schr\"{o}dinger equation for the standard Woods-Saxon potential with
zero angular momentum has no bound states. According to Eq.(3.16)
the energy eigenvalues depend on the depth of the potential $V_{0}$,
the width of the potential $R_{0}$, the thickness $a$ surface and
$D$ parameter. Any energy eigenvalue must not be less than $-V_{0}$,
i.e., $-V_{0}<E<0$. If constraints imposed on $n_{r}$, $V_{0}$ and
$E$ satisfied, the bound states appear. From Eq.(3.15) is seen that
the potential depth increases when the parameter $a$ increases, but
the parameter $R_{0}$ is decreasing for given $l$ quantum number and
vice versa. Therefore, one can say that the bound states exist
within this potential.

In addition, we have seen that there are some restrictions on the
potential parameters for the bound state solutions within the
framework of quantum mechanics. Hence, when the values of the
parameters $n_{r}$, $V_{0}$ and energy eigenvalues $E$ satisfy the
conditions in Eqs.(3.14), (3.15) and $-V_{0}<E<0$ respectively, we
obtain the bound states. We also point out that the exact results
obtained for the standard Woods-Saxon potential may have some
interesting applications for studying different quantum mechanical
and nuclear scattering problems. Consequently, the found wave
functions are physical ones.

Now, we are going to determine the radial eigenfunctions of this
potential. Having substituted $\pi (z)$ and $\sigma (z)$ into
Eq.(A.4) and then solving first-order differential equation, one can
find the finite function $\Phi (z)$ in the interval $(0,\, \, 1)$

\begin{equation}
\Phi (z)=z^{\varepsilon } (1-z)^{\sqrt{\varepsilon ^{2} -\beta ^{2}
+\gamma ^{2} } } .
\end{equation}

It is easy to find the second part of the wave function from the
definition of weight function

\begin{equation}
\rho (z)=z^{2\varepsilon } (1-z)^{2\sqrt{\varepsilon ^{2} -\beta
^{2} +\gamma ^{2} } }
\end{equation}
and substituting into Rodrigues relation Eq.(A.5), we get

\begin{equation}
y_{n_{r} } (z)=B_{n_{r} } z^{-2\varepsilon }
(1-z)^{-2\sqrt{\varepsilon ^{2} -\beta ^{2} +\gamma ^{2} } }
\frac{d^{n_{r} } }{dz^{n_{r} } } \left[z^{n_{r} +2\varepsilon }
(1-z)^{n_{r} +2\sqrt{\varepsilon ^{2} -\beta ^{2} +\gamma ^{2} } }
\right],
\end{equation}
where $B_{n_{r} } $ is the normalization constant and its value is
$\frac{1}{n_{r} !} $ ~\cite {Bateman}. Then, $y_{n_{r} } $ is given
by the Jacobi polynomials

\[y_{n_{r} } (z)=P_{n_{r} }^{(2\varepsilon \, ,\, \, 2\sqrt{\varepsilon ^{2} -\beta ^{2} +\gamma ^{2} } )} (1-2z),\]
where
\[P_{n}^{(\alpha \, ,\, \beta )} (1-2z)=\frac{1}{n!} z^{-\alpha } (1-z)^{-\beta } \frac{d^{n} }{dz^{n} }
\left[z^{n+\alpha } (1-z)^{n+\beta } \right].\]

The corresponding $u_{n_{r}l}(z)$ radial wave functions are found
as:

\begin{equation}
u_{n_{r} l} (z)=C_{n_{r} l} z^{\varepsilon }
(1-z)^{\sqrt{\varepsilon ^{2} -\beta ^{2} +\gamma ^{2} } } P_{n_{r}
}^{(2\varepsilon \, ,\, \, 2\sqrt{\varepsilon ^{2} -\beta ^{2}
+\gamma ^{2} } )} (1-2z),
\end{equation}
where $C_{n_{r}l}$ is the normalization constant determined by using
the following orthogonality relation:

\begin{equation}
\int _{0}^{\infty }\left|R_{n_{r}l} (r)\right|^{2} r^{D-1}  dr=\int
_{0}^{\infty }\left|u_{n_{r}l} (r)\right|^{2} dr=a\int _{0}^{1
}\frac{\left|u_{n_{r}l} (z)\right|^{2} }{z(1-z)}  dz=1 .
\end{equation}

\section{\bf Solution of the $D$-dimensional Schr\"{o}dinger equation by Supersymmetric quantum mechanics method}

According to SYSYQM, the eigenfunction of ground state $u_{0} (r)$
in Eq.(2.21) is a form as below

\begin{equation}
u_{0} (r)=N\exp \left(-\frac{\sqrt{2\mu } }{\hbar } \int W(r)dr
\right),
\end{equation}
where $N$ is normalized constant and $W(r)$ is superpotential. The
connection between the supersymmetric partner potentials $V_{1}(r)$
and $V_{2} (r)$ of the superpotential $W(r)$ is as follows ~\cite
{Cooper}:

\begin{equation}
V_{1} (r)=W^{2} (r)-\frac{\hbar }{\sqrt{2\mu } } W'(r)+E\, \, ,\, \,
\, \, V_{2} (r)=W^{2} (r)+\frac{\hbar }{\sqrt{2\mu } } W'(r)+E .
\end{equation}

The particular solution of the Riccati equation Eq.(4.2) searches
the following form:

\begin{equation}
W(r)=-\frac{\hbar }{\sqrt{2\mu } }
\left(A+\frac{B}{1+e^{\frac{r-R_{0} }{a} } } \right),
\end{equation}
where $A$ and $B$ are unknown constants. Since $V_{1}
(r)=\tilde{V}_{eff} (r)$, having inserted the relations Eqs.(2.18)
and (4.3) into the expression Eq.(4.2), and from comparison of
compatible quantities in  the left and right sides of the equation,
we find the following relations:

\begin{equation}
A^{2} =-\frac{2\mu }{\hbar ^{2}}(E_{0} -K_{0} ) \, ,\, \, \, \,
2AB-\frac{B}{a} =-\frac{2\mu K_{1}}{\hbar ^{2}} \, ,\, \, \, \,
B^{2} +\frac{B}{a}=\frac{2\mu K_{2}}{\hbar ^{2}} \, .
\end{equation}
If we use Eqs.(2.23) for  $A$ and $B$ parameters, Eqs.(4.4) are as
follows:

\begin{equation}
A^{2} =\frac{\varepsilon ^{2}}{a^{2} } \, ,\, \, \, \,
2AB-\frac{B}{a} =-\frac{\beta ^{2}}{a^{2} } \, ,\, \, \, \, B^{2}
+\frac{B}{a} =\frac{\gamma ^{2}}{a^{2}} \, .
\end{equation}
After inserting Eq.(4.3) into Eq.(4.1) and solving the integral, the
eigenfunction for ground state is obtained as

\begin{equation}
u_{0} (r)=N\, e^{Ar} \left(1+e^{-\frac{r-R_{0} }{a}} \right)^{-aB} .
\end{equation}
$A$ must be less than zero, and $B$  must be greater than zero for
the radial $u_{0}(r)$ wave function satisfy the boundary conditions
$u_{0}(0)=0$ and $u_{0} (\infty )=0$. Under this circumstance,
Eqs.(4.5) are as follows:

\begin{equation}
A=\frac{1}{2a} -\frac{\beta ^{2} }{a\left(\sqrt{1+4\gamma ^{2} }
-1\right)} \, ,
\end{equation}

\begin{equation}
B=\frac{\sqrt{1+4\gamma ^{2} } -1}{2a} ,
\end{equation}

\begin{equation}
E_{0}^{(D)} =K_{0} -\frac{\hbar ^{2} }{2\mu} \left[\frac{1}{2a}
-\frac{\beta ^{2} }{a\left(\sqrt{1+4\gamma ^{2} } -1\right)}
\right]^{2} .
\end{equation}

When $r\to \infty $, the chosen superpotential $W(\, r)$ is $W(\,
r)\to -\frac{\hbar A}{\sqrt{2\mu }}$. Having inserted Eq.(4.3) into
Eq.(4.2), for supersymmetric partner potentials, we obtain:

\begin{equation}
V_{1} (r)=\frac{\hbar ^{2} }{2\mu } \left[A^{2} +\frac{B^{2}
+\frac{B}{a} }{\left(1+e^{\frac{r-R_{0} }{a} } \right)^{2} }
+\frac{2AB-\frac{B}{a} }{1+e^{\frac{r-R_{0} }{a} } } \right]
\end{equation}
and
\begin{equation}
V_{2} (r)=\frac{\hbar ^{2} }{2\mu } \left[A^{2} +\frac{B^{2}
-\frac{B}{a} }{\left(1+e^{\frac{r-R_{0} }{a} } \right)^{2} }
+\frac{2AB+\frac{B}{a} }{1+e^{\frac{r-R_{0} }{a} } } \right] .
\end{equation}

If we add side-by-side the second equation of Eqs.(4.5) to third
equation of Eqs.(4.5), we obtain:

\begin{equation}
\, 2AB+B^{2} =\frac{\gamma ^{2} -\beta ^{2} }{a^{2} }
\end{equation}
from here

\begin{equation}
\, A=\frac{\gamma ^{2} -\beta ^{2} }{2a^{2} B} -\frac{B}{2} .
\end{equation}

Two partner potentials $V_{1} (\, r)$ and $V_{2} (\, r)$ which
differ from each other with additive constants and have the same
functional form are called invariant potentials
~\cite{Gendenshtein1, Gendenshtein2}. Thus, for the partner
potentials $V_{1} (\, r)$ and $V_{2} (\, r)$ given with Eqs.(4.10)
and (4.11), the invariant forms are:
\begin{equation}
R(B_{1} )=V_{2} (B,\, r)-V_{1} (B_{1} ,r)=-\frac{\hbar ^{2} }{2\mu }
\left[\left(\frac{\gamma ^{2} -\beta ^{2} }{2a^{2}
\left(B-\frac{1}{a} \right)} -\frac{B-\frac{1}{a} }{2} \right)^{2}
-\left(\frac{\gamma ^{2} -\beta ^{2} }{2a^{2} B} -\frac{B}{2}
\right)^{2} \right] ,
\end{equation}

\begin{equation}
\begin{array}{l} R(B_{i} )=V_{2} \left[B-\frac{i-1}{a} \, ,\, r\right]-V_{1}
\left[B-\frac{i}{a} \, ,\, r\right] = \\ -\frac{\hbar ^{2} }{2\mu }
\left[\left(\frac{\gamma ^{2} -\beta ^{2} }{2a^{2}
\left(B-\frac{i}{a} \right)} -\frac{B-\frac{i}{a} }{2} \right)^{2}
-\left(\frac{\gamma ^{2} -\beta ^{2} }{2a^{2} \left(B-\frac{i-1}{a}
\right)} -\frac{B-\frac{i-1}{a} }{2} \right)^{2} \right] .
\end{array}\end{equation}

If we continue this procedure and make the substitution
$\,B_{n_{r}}=B_{n_{r}-1} -\frac{1}{a} =B-\frac{n_{r}}{a}$ at every
step until $\, B_{n_{r}} \ge 0$, the whole discrete spectrum of
Hamiltonian $\, H_{-}(B)$:

\[E_{n_{r}l}^{(D)} =E_{0}^{(D)} +\sum _{i=1}^{n_{r}}R(B_{i} )\, ,\]

\begin{equation}
\begin{array}{l} E_{n_{r}l}^{(D)} =K_{0} -\frac{\hbar ^{2} }{2\mu } \left[\left(\frac{\gamma ^{2} -\beta ^{2} }{2a^{2}
\left(B-\frac{n_{r}}{a} \right)} -\frac{B-\frac{n_{r}}{a} }{2}
\right)^{2} -\left(\frac{\gamma ^{2} -\beta ^{2} }{2a^{2}
\left(B-\frac{n_{r}-1}{a} \right)} -\frac{B-\frac{n_{r}-1}{a} }{2}
\right)^{2} +\left(\frac{\gamma ^{2} -\beta ^{2} }{2a^{2}
\left(B-\frac{n_{r}-1}{a} \right)} -\frac{B-\frac{n_{r}-1}{a} }{2}
\right)^{2} -\right. \\ {\, \, -\left(\frac{\gamma ^{2} -\beta ^{2}
}{2a^{2} \left(B-\frac{n_{r}-2}{a} \right)}
-\frac{B-\frac{n_{r}-2}{a} }{2} \right)^{2} +\ldots
+\left(\frac{\gamma ^{2} -\beta ^{2} }{2a^{2} \left(B-\frac{2}{a}
\right)} -\frac{B-\frac{2}{a} }{2} \right)^{2} -\left(\frac{\gamma
^{2} -\beta ^{2} }{2a^{2} \left(B-\frac{1}{a} \right)}
-\frac{B-\frac{1}{a} }{2} \right)^{2} +} \\ {\, \, \left. +\,
\left(\frac{\gamma ^{2} -\beta ^{2} }{2a^{2} \left(B-\frac{1}{a}
\right)} -\frac{B-\frac{1}{a} }{2} \right)^{2} -\left(\frac{\gamma
^{2} -\beta ^{2} }{2a^{2} B} -\frac{B}{2} \right)^{2}
+\left(\frac{1}{2a} -\frac{\beta ^{2}
}{a\left(\sqrt{1+4\gamma ^{2} } -1\right)} \right)^{2} \right]=} \\
{\, \, =K_{0} -\frac{\hbar ^{2} }{2\mu } \left[\frac{\gamma ^{2}
-\beta ^{2} }{2a^{2} \left(B-\frac{n_{r}}{a} \right)}
-\frac{B-\frac{n_{r}}{a} }{2} \right]^{2} =K_{0} -\frac{\hbar ^{2}
}{2\mu \, a^{2} } \left[\frac{\beta ^{2} -\gamma ^{2}
}{\sqrt{1+4\gamma ^{2} } -2n_{r}-1} +\frac{\sqrt{1+4\gamma ^{2} }
-2n_{r}-1}{4} \right]^{2} } .
\end{array}
\end{equation}

Thereby, if we consider the parameter $\varepsilon ,\, \, \beta ,\,
\gamma $ into Eq.(4.16), the obtained expression for energy
eigenvalue in $l$-state will be same with expression Eq.(3.16) which
was obtained by NU method. When $D=3$, there are no any bound states
in system for $l=0$, because the inequalities $A<0\, ,\, B>0$ are
not satisfied. As a result ,  $n_{r}$ is less than zero and the
calculated energy eigenvalues do not satisfy the inequality
$-V_{0}<E<0$. It should be noted that the same  conditions for
$n_{r}$ and $V_{0} $ in Eq.(3.14) and Eq.(3.15) obtained by NU
method are also determined from the following inequalities $B>0, \,
A<0$. When $D>3$, there are bound states in system. It is seen from
Eq.(3.16), the energy eigenvalue depends on the depth $V_{0}$ of the
potential, the width $R_{0} $ potential, the thickness $a$ surface,
and $D$ parameter. Thus, the determined conditions for $n_{r}$,
$V_{0}$ and $E$, i.e., if the inequities $A<0\, , B>0$ and
$-V_{0}<E<0$ are satisfied respectively, there are the bound states
in the system, and the energy spectrum of these states is limited
number. Based on Eqs.(B.15) and (B.17), the obtained result of
radial Schr\"{o}dinger equation by using the Eq.(4.6) of the ground
state eigenfunction is exactly same with the result obtained by
using NU method.

\section{\bf Results and Discussion}

In this chapter, in order to analyze the present qualitative
findings, the single particle energy levels, the effective
potentials and normalized wave functions of neutron moving under the
average potential field of the $^{56} Fe$ nucleus are calculated for
various $n_{r}$ and $l$ quantum numbers by using the empirical
values $r_{0}=1.285\, \, fm$ and $a=0.65\, \, fm$ taken from
Ref.~\cite {Perey}. Under these certain circumstances, the potential
depth of mentioned potential is $V_{0} =(40.5+0.13A)\, MeV=47.78\,
\, MeV$, and the radius of the nucleus is $R_{0} =r_{0}
A^{{\tfrac{1}{3}} } =4.9162\, \, fm$. Here $A$ is the atomic mass
number of $^{56} Fe$ nucleus. The reduced mass consists of neutron
mass $m_{n}=1.00866\, \, u$ and $^{56} Fe$ core mass with is
$m_{A}=56\, \, u$, and its value is $\mu =\frac{m_{A} \cdot
m_{n}}{m_{A} +m_{n}}=0.990814\, \, u$.

Calculated energies of the bound states and normalized wavefunction
for $D=3$ in different values of $n_{r}$ and $l$ are presented in
Table 1 indicate: when $1\leq l\leq 4$ and $n_{r}=0$, there are
bound states in system. However, when $1\leq l\leq 4$ in $n_{r}\geq
1$ and $5\leq l\leq 7$ in $n_{r}=0$ the energy values $E_{n_{r}l}$
are not satisfied the inequality $-V_{0}\leq E<0$, i.e., these
findings cannot be considered physically and are only the
mathematical results. Moreover, it is clearly seen from Fig.1 that
when $l=8$, the effective potential $V_{eff}(r)$ decreases
monotonically, so there is no solution in transcendental equation
i.e., there is neither physical nor mathematical result. Hence,
there are no bound states in the system for the quantum numbers
$5\leq l \leq 7$. Furthermore, it can be clearly seen from Table I
and Fig.1 that when the value of quantum number $l$ increases, the
value of $r_{l}$ continues be closer the width of the nuclear radius
$R_{0}$, and when $l=7$, the value of $r_{l}$ is greater than the
width of the nuclear radius $R_{0}$. Namely, the reason why there
are no bound states in system for the quantum numbers $5\leq l \leq
7$ could be due to the centrifugal potential expanding in the series
around the surface of the nucleus. This fact is also confirmed by
the result of Refs.~\cite {Badalov1,Badalov2,Badalov3,Badalov4}
i.e., there are no bound states of system near the surface of the
nucleus. In addition, the general behavior of normalization wave
function for $1\leq l \leq 4$ by a comparison with it for $5\leq l
\leq 7$ as shown in Fig.2 is very different. This alteration
happened after $l=4$ which can be also related to the lack of bound
states in system for $5\leq l \leq 7$.

When  $D=4$, the calculated energies of the bound states and
normalized wavefunction in the different $n_{r}$ and $l$ values are
presented in Table 2. It is seen from Table 2, there are bound
states in system for $n_{r}=0$ and $1\leq l\leq 3$. Hence,
Eqs.(3.14) and (3.15) are satisfied for not only $l=0$ and $4\leq
l\leq 6$ in $n_{r}\geq 0$, but also $1\leq l\leq 3$  in $n_{r}\geq
1$. However, the energy eigenvalues do not satisfy the inequality
$-V_{0}\leq E<0$. It means that there are not the bound states in
system for the mentioned $n_{r}$ radial and $l$ orbital quantum
numbers. It should be noted that when $l\geq7$ , there is no
solution in transcendental equation.

It seen from Table 1 and Table 2, the energy of the bound states
increases with increasing of $D$ in the fixed same values of $n_{r}$
and $l$, i.e., $E^{(4)}_{01}>E^{(3)}_{01},
E^{(4)}_{02}>E^{(3)}_{02}, E^{(4)}_{03}>E^{(3)}_{03}$ . It means
that the repulsive force appears in system owing to the additional
centrifugal potential $V_{l}(r)=\frac{\hbar ^{2}
\tilde{l}(\tilde{l}+1)}{2\mu r^{2} }$. Therefore, in order to
compensate this potential the energy of the bound state must
increase \cite{Greiner,Ballentine}. Note that when $D=3$ and $D=4$,
there are not the bound states in system for $n_{r}=0, l=0$.
Nevertheless, when $D=5$, there are the bound states in system for
$n_{r}=0, l=0$ and its enegies is $E^{(5)}_{00}=-42.8980454 MeV$.
Thus, the reason why there are not the bound states in system for
$n_{r}=0, l=0$ when $D=3$ and $D=4$ is related with the standard
Woods-Saxon potential cannot describe the system fully. As a way out
of this, the modified version of the standard Woods-Saxon potential
such as the generalized Woods-Saxon potential and the spin and
pseudospin symmetries in the standard Woods-Saxon potential can be
utilized for solving the problem. Spin and pseudospin symmetries are
symmetries of the Dirak Hamiltonian. Thus, pseudospin symmetry was
discussed firstly in non-relativistic framework ~\cite{Hecht,Arima},
then in relativistic mean field theory. Comprehensive discussed in
the Refs. ~\cite{Liang1,Ginocchio1,Meng1,Meng2,Chen,Zhou,Ginocchio2,
Meng3,Liang2,Liang3} spin and pseudospin symmetries will be utilized
for my further studies.

\section{\bf Conclusion}
To conclude, an analytical study of the D-dimensional space
Schr\"{o}dinger equation have been performed for Woods-Saxon
potential using the improved approximation scheme to the centrifugal
term for arbitrary $l$-states. There the energy eigenvalues of the
bound states and corresponding eigenfunctions have been analytically
found via both of NU and SUSYQM methods within the Pekeris
approximation. The same expressions were obtained for the energy
eigenvalues, and the expression of radial wavefunctions transformed
each other was also shown by using these methods. The energy
eigenvalues depending on $V_{0} \, ,\, \, R_{0} \, ,\, \, a$ and $D$
parameters have a finite number energy spectrum for standard
Wood-Saxon potential, so it puts some restrictions on the potential
parameters during the solution of related cases within the framework
of quantum mechanics. In this way, if the potential parameters
$V_{0}$, $n_{r}$ and energy eigenvalues $E$ satisfy the conditions
in Eqs.(3.14), (3.15) and $-V_{0}<E<0$ respectively, it means there
are bound states in system. It should be noted that the same
limiting conditions were obtained for $V_{0} $ and $n_{r}$ thanks to
both methods. Since there is the practical interest for the energy
spectrum in various potentials, investigating the features of
eigenvalues is very important and actual with regard to arbitrary
parameter of system. For illustration, the bound states energies of
$\,\,^{56} Fe\,\,$ nucleus have been calculated and analyzed for
some $l$ and $n_{r}$ values. The qualitative results of this study
are expected to enable new possibilities for pure theoretical and
experimental physicists, because the results are exact and more
general.

\newpage
\begin{table}[h]
\begin{center}
\begin{tabular}{|c|c|c|c|c|c|}\hline
$n_{r} $ & $l$ & $r_{l} ,\;fm $ & $V_{eff,min} , MeV$ & $E_{n_{r}l} , MeV$ & $u_{n_{r}l}(z) $ \\
\hline

0 & 1 & 2.95578498158 & -40.71121848 & -42.8980494 &
$7.419162631z^{3.913357119}(1-z)^{0.2835207487} $ \\ \hline

1 & 1 & 2.95578498158 & -40.71121848 & -164.0083691 & $ Unbound $
\\ \hline

0 & 2 & 3.43967490298 & -32.59671725 & -30.9674480&
$4.630848265z^{2.521449526}(1-z)^{0.1881862542} $ \\ \hline

1 & 2 & 3.43967490298 & -32.59671725 & -174.5240650& $ Unbound $ \\
\hline

0 &3& 3.78599536866 & -22.94860534 & -18.3133413 &
$3.248777890z^{1.7924708185}(1-z)^{0.09698158616} $ \\ \hline

1 &3& 3.78599536866 & -22.94860534 & -209.1611062 & $ Unbound $ \\
\hline

0 & 4 & 4.07888427247 & -12.05092239 & -5.16198171 &
$2.366451293z^{1.293697429}(1-z)^{0.003675679733} $ \\ \hline

1 & 4 & 4.07888427247 & -12.05092239 & -385.5364626 & $ Unbound $ \\
\hline

0 & 5 & 4.35562101985 & -0.18380311 & 8.03190791 &
$2.403325566z^{0.895619899}(1-z)^{0.08910171673} $  \\ \hline

0 & 6 & 4.65152782641 & 12.32885320 & 20.44480441 &
$2.284361438z^{0.5317537204}(1-z)^{0.1866465558} $ \\ \hline

0 & 7 & 5.07501734690 & 24.95664294 & 20.79588752 &
$2.835605734z^{0.4259909093}(1-z)^{0.4933508522} $ \\ \hline
\end{tabular}
\end{center}

\caption{Calculated energies of the bound states and normalized
wavefunction for $V_{0} =47.78 MeV$, $R_{0} =4.9162 fm$, $a=0.65
fm$, $D=3$ in different values of $n_{r}$ and $l$.} \label{table 1}
\end{table}

\newpage
\begin{table}[h]
\begin{center}
\begin{tabular}{|c|c|c|c|c|c|}\hline
$n_{r} $ & $l$ & $r_{l} ,\;fm $ & $V_{eff,min} , MeV$ & $E_{n_{r}l} , MeV$ & $u_{n_{r}l}(z) $ \\
\hline

0 & 0 & 2.56619312728 & -44.12161049 & -48.730119 & $ Unbound $ \\
\hline

1 & 0 & 2.56619312728 & -44.12161049 & -161.448867 & $ Unbound $ \\
\hline

0 & 1 & 3.22513008574 & -36.86366841 & -37.0225964 &
$5.722788228z^{3.074200941}(1-z)^{0.2347858502} $ \\ \hline

1 & 1 & 3.22513008574 & -36.86366841 & -167.8844225 & $ Unbound $ \\
\hline

0 & 2 & 3.62275083960 & -27.94626879 & -24.7231461&
$3.847953167z^{2.114265381}(1-z)^{0.1424715417} $ \\ \hline

1 & 2 & 3.62275083960 & -27.94626879 & -186.4599846& $ Unbound $ \\
\hline

0 &3& 3.93641236772 & -17.63870684 & -11.7764513 &
$2.769256517z^{1.525007303}(1-z)^{0.05127042688} $ \\
\hline

1 &3& 3.93641236772 & -17.63870684 & -257.2474288 &
$ Unbound $ \\
\hline

0 & 4 & 4.21735772895 & -6.22020931 & 1.46781624 &
$ 2.382992171z^{1.0867462}(1-z)^{0.04134934644} $ \\
\hline

1 & 4 & 4.21735772895 & -6.22020931 & -1026.704467 &
$ Unbound $ \\
\hline

0 & 5 & 4.49813167590 & 6.01629080 & 14.41696183 &
$3.605399251z^{0.7131560907}(1-z)^{0.5085916097} $  \\
\hline

0 & 6 & 4.82904585786 & 18.68343922 & 25.77233519 &
$1.855453960z^{0.3396915672}(1-z)^{0.1153903608} $ \\
\hline

\end{tabular}
\end{center}

\caption{Calculated energies of the bound states and normalized
wavefunction for $V_{0} =47.78 MeV$, $R_{0} =4.9162 fm$, $a=0.65
fm$, $D=4$ in different values of $n_{r}$ and $l$.} \label{table 2}
\end{table}

\newpage


\begin{figure}[!hbt]
\vskip 1.2cm\epsfxsize 9.8cm \centerline{\epsfbox{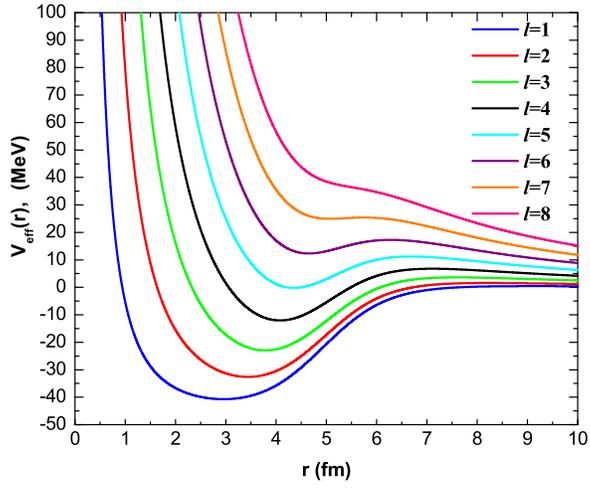}}

\vskip-0.2cm \caption{The effective potential $V_{eff} (r)$  as a
function of the internuclear  $r$- distance and several $l$ quantum
numbers  for $V_{0} =47.78 MeV$, $R_{0} =4.9162 fm$, $a=0.65 fm$,
$D=3$.} \label{Fig1}
\end{figure}

\begin{figure}[!hbt]
\vskip 1.2cm\epsfxsize 9.8cm \centerline { \epsfbox{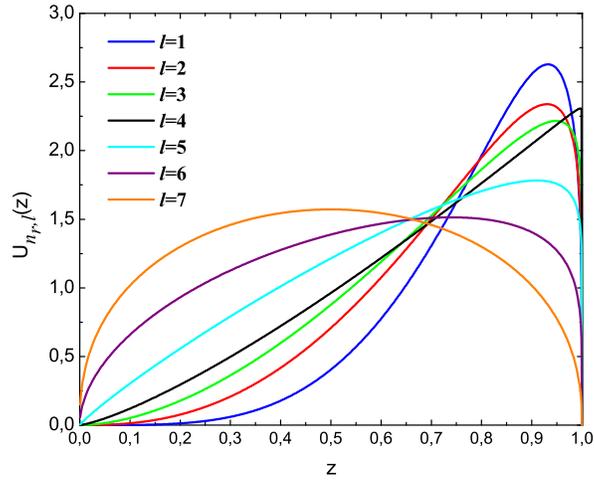}}
\vskip-0.2cm \caption{The normalized wave function as a function of
$z$ and several quantum numbers $l$ for $V_{0} =47.78 MeV$, $R_{0}
=4.9162 fm$, $a=0.65 fm$, $D=3$, $n_{r}=0$.} \label{Fig2}
\end{figure}

\newpage

\section*{\bf {APPENDIX A. Nikiforov-Uvarov method}}

The NU method is based on the solutions of general second order
linear equations with special orthogonal functions. It has been
extensively used to solve the non-relativistic Schr\"{o}dinger
equation and other Schr\"{o}dinger-like equations. The
one-dimensional (1D) Schr\"{o}dinger equation or similar
second-order differential equations can be written with NU method in
the following form ~\cite{Nikiforov}:

\qquad\qquad\qquad\qquad\qquad\qquad$\psi ''(z)+\frac{\tilde{\tau
}(z)}{\sigma (z)} \psi'(z)+\frac{\tilde{\sigma }(z)}{\sigma ^{2}(z)}
\psi(z)=0 \, ,$ \qquad\qquad\qquad\qquad\qquad (A.1) where $\sigma
(z)$ and $\tilde{\sigma }(z)$ are polynomials at most second-degree,
and $\tilde{\tau }(z)$ is a first-degree polynomial.

By using in Eq.(A.1) the transformation

\qquad\qquad\qquad\qquad\qquad\qquad\qquad\qquad $\psi
(z)=\Phi(z)y(z)$ \qquad\qquad\qquad\qquad\qquad\qquad\,\, (A.2) one
reduces it to the hypergeometric-type equation

\qquad\qquad\qquad\qquad\qquad\qquad $\sigma (z)y''+\tau
(z)y'+\lambda y=0 \,.$\qquad\qquad\qquad\qquad\qquad\qquad (A.3) The
function $\Phi (z)$ is defined as the logarithmic derivative
~\cite{Nikiforov}

\qquad\qquad\qquad\qquad\qquad\qquad\qquad\qquad $\frac{\Phi
'(z)}{\Phi (z)} =\frac{\pi (z)}{\sigma (z)} \, ,$
\qquad\qquad\qquad\qquad\qquad\qquad\qquad\,\, (A.4) where $\sigma
(z)$ is at most the first-degree polynomial.

The another part of $\psi (z)$, namely $y(z)$ is the
hypergeometric-type function, that for fixed $n$ is given by the
Rodriguez relation

\qquad\qquad\qquad\qquad\qquad\qquad $ y_{n} (z)=\frac{B_{n} }{\rho
(z)} \frac{d^{n} }{dz^{n} } [\sigma ^{n} (z)\rho (z)] \, ,$
\qquad\qquad\qquad\qquad\qquad\qquad (A.5) where $B_{n}$ is the
normalization constant and the weight function $\rho (z)$ must
satisfy the condition ~\cite{Nikiforov}

\qquad\qquad\qquad\qquad\qquad\qquad\qquad \,\,\,$\frac{d}{dz}
(\sigma (z)\rho (z))=\tau (z)\rho (z) ,$
\qquad\qquad\qquad\qquad\,\,\, (A.6) with $\tau (z)=\tilde{\tau
}(z)+2\pi (z) \, .$

For accomplishment of the conditions imposed on function $\rho (z)$
the classical orthogonal polynomials, it is necessary, that
polynomial $\tau (z)$ becomes equal to zero in some point of an
interval $(a,b)$ and derivative of this polynomial for this interval
at $\sigma (z)>0$ will be negative, i.e. $\tau '(z)<0 \, $.

The function $\pi (z)$ and the parameter $\lambda $ required for
this method are defined as follows ~\cite{Nikiforov}

\qquad\qquad\qquad\qquad\qquad\qquad $\pi (z)=\frac{\sigma
'-\tilde{\tau }}{2} \pm \sqrt{\left(\frac{\sigma '-\tilde{\tau }}{2}
\right)^{2} -\tilde{\sigma }+k\sigma} \, ,$ \qquad\qquad\qquad\qquad
\,\,  (A.7)

\qquad\qquad\qquad\qquad\qquad\qquad\qquad\qquad $\lambda =k+\pi
'(z) \,.$ \qquad\qquad\qquad\qquad\qquad\qquad\,\,\,\,\,\,\,\, (A.8)

On the other hand, in order to find the value of $k$, the expression
under the square root must be the square of a polynomial. This is
possible only if its discriminant is zero. Thus, the new eigenvalue
equation for the Schr\"{o}dinger equation becomes ~\cite{Nikiforov}

\qquad\qquad\qquad\qquad\qquad $\lambda =\lambda _{n} =-n\tau
'-\frac{n(n-1)}{2} \sigma ''\, \, ,\, \, \, (n=0,\, 1,\, 2,\, ...\,
)\,.$ \qquad\qquad\qquad (A.9)
After the comparison of Eq.(A.8) with Eq.(A.9), we obtain the energy
eigenvalues.

\section*{\bf APPENDIX B. Supersymmetric Quantum Mechanics}

SUSYQM for $N=2$, we have two nilpotent operators, $Q$ and $Q^{+} $,
satisfying the following algebra:

\quad\quad\quad\quad\quad\quad\quad\quad $\{ Q,\, Q^{+} \}=H\, ,\,
\, \, \, \, \, \{ Q\, ,\, Q\}=\{ Q^{+} ,Q^{+} \} =0 \, ,$
\quad\quad\quad\quad\quad\quad\quad\quad\quad (B.1) where $H$ is the
supersymmetric Hamiltonian, $Q=\left(\begin{array}{cc} {0} & {0} \\
{A^{-} } & {0}
\end{array}\right)$ and $Q^{+} =\left(\begin{array}{cc} {0} & {A^{+}
} \\ {0} & {0} \end{array}\right)$ are the operators of
supercharges, $A^{-} $ is bosonic operators and $A^{+}$ is its
adjoint. The supersymmetric $H$ Hamiltonian is given by
~\cite{Cooper}:

\quad\quad\quad\quad\quad\quad\quad\quad$ H=\left(\begin{array}{cc}
{A^{+} A^{-} } & {0} \\ {0} & {A^{-} A^{+}
} \end{array}\right)\, =\left(\begin{array}{cc} {H_{-} } & {0} \\
{0} & {H_{+} }
\end{array}\right) ,$\quad\quad\quad\quad\quad\quad\quad\quad\quad\quad
(B.2) where $H_{-} $ and $H_{+} $ are called supersymmetric partner
Hamiltonians. The supercharges $Q$ and $Q^{+} $ commute with SUSY
$H$ Hamiltonian: $[H\, ,\, Q]=[H,Q^{+} ]=0$.

If the ground state energy of a Hamiltonian $H$ is zero (i.e. $E_{0}
=0$), it can always be written in a factorable form as a product of
a pair of linear differential operators. That is why, one has from
the Schr\"{o}dinger equation that the ground state wave function
$\psi _{o} (x)$ obeys

\quad\quad\quad\quad\quad\quad\quad\quad\quad\quad $ H\psi _{o}
(x)=-\frac{\hbar ^{2} }{2m} \frac{d^{2} \psi _{o} }{dx^{2} }
+V(x)\psi _{o} (x)=0 $
\quad\quad\quad\quad\quad\quad\quad\quad\quad\quad     (B.3)
so that

\quad\quad\quad\quad\quad\quad\quad\quad\quad\quad\quad\quad\quad\quad\quad
$ V(x)=\frac{\hbar ^{2} }{2m} \frac{\psi ''_{o} (x)}{\psi _{o}
(x)}\,.$
\quad\quad\quad\quad\quad\quad\quad\quad\quad\quad\quad\quad\quad\quad
(B.4)

This allows a global reconstruction of the potential $V(x)$ from the
knowledge of its ground state wave function which has no nodes. Once
we realize this, it is now very simple to factorize the Hamiltonian
using the following ansatz ~\cite{Cooper}:

\quad\quad\quad\quad\quad\quad\quad\quad\quad\quad\quad\quad $ H_{-}
=-\frac{\hbar ^{2} }{2m} \frac{d^{2} }{dx^{2} } +V(x)=A^{+} A^{-}\,
,$ \quad\quad\quad\quad\quad\quad\quad\quad\quad\quad (B.5) where

\quad\quad\quad\quad\quad\quad\quad\quad\quad $ A^{-} =\frac{\hbar
}{\sqrt{2m} } \frac{d}{dx} +W(x)\,, \,  A^{+} =-\frac{\hbar
}{\sqrt{2m} } \frac{d}{dx} +W(x)\, .$
\quad\quad\quad\quad\quad\quad\quad (B.6)

By factorizing procedure of the Hamiltonian, the Riccati equation
for Superpotential  is obtained:

\quad\quad\quad\quad\quad\quad\quad\quad\quad\quad\quad $ V_{-}
(x)=W^{2} (x)-\frac{\hbar }{\sqrt{2m} } W'(x) \,.$
\quad\quad\quad\quad\quad\quad\quad\quad\quad\quad\quad\quad (B.7)

The solution for $W(x)\, $ in terms of the ground state wave
function is

\quad\quad\quad\quad\quad\quad\quad\quad\quad\quad\quad\quad\quad\quad\quad\quad\,\,
$ W(x)=-\frac{\hbar}{\sqrt{2m}} \frac{\psi '_{o} (x)}{\psi _{o} (x)}
\,.$ \quad\quad\quad\quad\quad\quad\quad\quad\quad\quad \,\,(B.8)

This solution is obtained by recognizing that once we satisfy $A^{-}
\psi _{0} (x)=0$, we automatically have a solution to $H\psi _{0}
=A^{+} A^{-} \psi _{0} =0 \,.$

The next step in constructing the SUSY theory related to the
original Hamiltonian $H_{-} $ is to define the operator $H_{+}
=A^{-} A^{+} $ obtained by reversing the order of $A^{-} $ and
$A^{+} $. A little simplification shows that the operator $H_{+} $
is in fact a Hamiltonian corresponding to a new potential $V_{+}
(x)$.

\quad\quad\quad\quad\quad\quad\quad\quad $ H_{+} =-\frac{\hbar ^{2}
}{2m} \frac{d^{2} }{dx^{2} } +V_{+} (x)\, \, \, ,\, \, \, \, V_{+}
(x)=W^{2} (x)+\frac{\hbar }{\sqrt{2m} } W'(x)\, .$
\quad\quad\quad\quad   (B.9)

The potentials $V_{-} (x)$ and $V_{+} (x)$ are known as
supersymmetric partner potentials. It is then clear that if the
ground state energy of a Hamiltonian $H_{1}$ is $E^{(1)}_{(0)}$ with
eigenfunction $\psi _{0}^{(1)}$ then in view of Eq.(B.5), it can
always be written in the form below as,

\quad\quad\quad\quad\quad\quad\quad\quad\quad $ H_{1} =-\frac{\hbar
^{2} }{2m} \frac{d^{2} }{dx^{2} } +V_{1} (x)=A^{+} A^{-}
+E_{0}^{(1)}\, ,$ \quad\quad\quad\quad\quad\quad\quad\quad\quad\quad
(B.10)
where

\quad\quad\quad\quad \,\,$\begin{array}{l} {A_{1}^{-} =\frac{\hbar
}{\sqrt{2m} } \frac{d}{dx} +W_{1} (x)\, ,\, \, A_{1}^{+}
=-\frac{\hbar }{\sqrt{2m} } \frac{d}{dx} +W_{1} (x),}
\\ V_{1} (x)=W_{1}^{2} (x)-\frac{\hbar }{\sqrt{2m} } W'_{1}
(x)+E_{0}^{(1)} , \quad W_{1} (x)=-\frac{\hbar}{\sqrt{2m}}\frac {d
\ln \psi_{0}^{(1)}}{dx} \,.\end{array}$ \quad\quad\quad\quad (B.11)

The SUSY partner Hamiltonian is then given by ~\cite{Cooper}

\quad\quad\quad\quad\quad\quad\quad\quad\quad\quad $ H_{2}
=A_{1}^{-} A_{1}^{+} +E_{0}^{(1)} =-\frac{\hbar ^{2} }{2m}
\frac{d^{2} }{dx^{2} } +V_{2} (x),$
\quad\quad\quad\quad\quad\quad\quad\quad\quad\,  (B.12) where

\, $ \, V_{2} (x)=W_{1}^{2} (x)+\frac{\hbar }{\sqrt{2m} } W'_{1}
(x)+E_{0}^{(1)} =V_{1} (x)+\frac{2\hbar }{\sqrt{2m} } W'_{1}
(x)=V_{1} (x)-\frac{\hbar^{2} }{m} \frac{d^{2} }{dx^{2} } (\ln \psi
_{0}^{(1)}).$ \quad (B.13)

From Eq.(B.12), the energy eigenvalues and eigenfunctions of the two
Hamiltonians $H_{1}$ and $H_{2}$ are related by ~\cite{Cooper}

\, $ E_{n}^{(2)} =E_{n+1}^{(1)} \, ,\, \, \, \, \psi
_{n}^{(2)}=[E_{n+1}^{(1)} -E_{0}^{(1)} ]^{-\frac{1}{2}}A_{^{1} }^{-}
\psi _{n+1}^{(1)}\, ,\, \, \, \psi _{n+1}^{(1)} =[E_{n}^{(2)}
-E_{0}^{(1)} ]^{-\frac{1}{2}}A_{^{1} }^{+} \psi _{n}^{(2)}\,.$
\quad\quad (B.14) Here $E_{n}^{(m)}$ is the energy level, where $n$
denotes the energy level and $(m)$ refers to the $m$'th Hamiltonian
$H_{m}$.

In this way, it is clear that if the original Hamiltonian $H_{1}$
has $p\ge 1$ bound states with eigenvalues $E_{n}^{(1)}$, and
eigenfunctions $\psi _{n}^{(1)}$ with $0<n<p$, then we can always
generate a hierarchy of $(p-1)$ Hamiltonians $H_{2} \, ,\, \, H_{3}
\, ,\, ...,\, H_{p}$ such that the $m$'th member of the hierarchy of
Hamiltonians $(H_{m})$ has the same eigenvalue spectrum as $H_{1}$
except that the first $(m-1)$ eigenvalues of $H_{1}$ are missing in
$H_{m}$ ~\cite{Cooper}:

\quad\quad\quad\quad\quad\quad\quad\quad\quad\quad\quad $ H_{m}
=A_{m}^{+} A_{m}^{-} +{\rm \; E}_{{\rm m-1}}^{{\rm (1)}}
=-\frac{\hbar ^{2} }{2m} \frac{d^{2} }{dx^{2} } +V_{m} (x) \,,$
\quad\quad\quad\quad\quad\quad (B.15) where

\quad $ A_{m}^{-} =\frac{\hbar }{\sqrt{2m} } \frac{d}{dx} +W_{m}
(x)\, ,\, \, \, \, W_{m} (x)=-\frac{\hbar }{\sqrt{2m} } \frac{d\ln
\psi _{0}^{(m)} }{dx} \, \, ,\, \, \, \, m=2\, ,\, 3\, ,\, \, 4,\,
\, \cdots ,\, \, p \,.$ \quad\quad\,\,\, (B.16)

One also has

\quad\quad\quad $
\begin{array}{l} E_{n}^{(m)}=E_{n+1}^{(m-1)}=\cdots =E_{n+m-1}^{(1)} \,, \\ \psi _{n}^{(m)} =[E_{n+m-1}^{(1)} -E_{m-2}^{(1)}
]^{-\frac{1}{2}} \cdots [E_{n+m-1}^{(1)}
-E_{0}^{(1)}]^{-\frac{1}{2}} A_{m-1}^{-} \cdots A_{1}^{-}
\psi^{(1)}_{n+m-1} \,, \\  V_{m}(x)=V_{1} (x)-\frac{\hbar^{2} }{m}
\frac{d^{2} }{dx^{2} } \ln (\psi _{0}^{(1)} \cdots \psi
_{0}^{(m-1)})\, .
\end{array}$\quad \,\,\, (B.17)
i.e., knowing all the eigenvalues and eigenfunctions of $H_{1}$ we
immediately know all the energy eigenvalues $E_{n}^{(1)}$ and
eigenfunctions $\psi _{n}^{(1)}$ of the hierarchy of $(p-1)$
Hamiltonians $H_{2} \, ,\, \, H_{3} \, ,\, ...,\, H_{p}.$

\end{document}